\documentclass[aps,prb,superscriptaddress,showpacs,twocolumn,10pt]{revtex4-1}

\usepackage{graphicx}
\usepackage{epstopdf}
\usepackage{amsmath}
\usepackage{amsfonts}
\usepackage{amssymb}
\usepackage{hyperref}
\usepackage{makecell}

\begin{document}

\title{Multichannel charge Kondo effect and non-Fermi liquid fixed points \\ in conventional and topological superconductor islands}

\author{Micha{\l} Papaj}
\author{Zheng Zhu}
\author{Liang Fu}
\affiliation{Department of Physics, Massachusetts Institute of Technology, Cambridge, Massachusetts 02139, USA}

\date{\today}

\begin{abstract}
We study multiterminal Majorana and conventional superconducting islands in the vicinity of the charge degeneracy point using bosonization and numerical renormalization group. Both models map to the multichannel charge Kondo problem, but for noninteracting normal leads they flow to different non-Fermi liquid fixed points at low temperatures. We compare and contrast both cases by numerically obtaining the full crossover to the low temperature regime and predict distinctive transport signatures. We attribute the differences between both types of islands to a crucial distinction of charge-$2e$ and charge-$e$ transfer in the conventional and topological case, respectively. In the conventional case, our results establish s-wave islands as a new platform to study the intermediate multichannel Kondo fixed point. In the topological setup the crossover temperature to non-Fermi liquid behavior is relatively high as it is proportional to level broadening and the transport results are not sensitive to channel coupling anisotropy, moving away from the charge degeneracy point or including a small Majorana hybridization, which makes our proposal experimentally feasible.

\end{abstract}


\maketitle

\section{Introduction.}

The prospect of robust quantum computation using Majorana zero modes \cite{KitaevFaulttolerantquantumcomputation2003, NayakNonAbeliananyonstopological2008, BeenakkerSearchMajoranaFermions2013a, AliceaNewdirectionspursuit2012a, SarmaMajoranazeromodes2015} sparked enormous experimental interest in the material platforms that enable direct observation and study of these topological quasiparticles. Among the leading platforms are proximitized semiconductor nanowires with spin-orbit coupling, which are predicted to become topological superconductors and host Majorana zero modes under external magnetic fields \cite{LutchynMajoranaFermionsTopological2010, OregHelicalLiquidsMajorana2010}. The immense experimental effort over the last several years has resulted in a significant improvement in material and device fabrication quality, and also helped to rekindle interest in mesoscopic superconductivity in semiconductor devices  \cite{MourikSignaturesMajoranaFermions2012, DasZerobiaspeakssplitting2012b, RokhinsonfractionalJosephsoneffect2012, ChurchillSuperconductornanowiredevicestunneling2013a, FinckAnomalousModulationZeroBias2013, AlbrechtExponentialprotectionzero2016a, KrogstrupEpitaxysemiconductorsuperconductor2015a, PlissardFormationelectronicproperties2013, ChangHardgapepitaxial2015, HigginbothamParitylifetimebound2015, GulBallisticMajoranananowire2018, DengMajoranaboundstate2016b, TaupinInAsNanowireEpitaxial2016, ShermanNormalsuperconductingtopological2017, ChenExperimentalphasediagram2017, GulHardSuperconductingGap2017, ZhangQuantizedMajoranaconductance2018}.

When considering the physics of mesoscopic conductors and superconductors it is crucial to take into account the Coulomb blockade effect that arises due to electron-electron interaction. Since the charging energy of the island depends quadratically on the number of electrons it contains, it is possible to use an external gate to tune two charge states of the island to be equal in energy. In conventional superconductors, where putting an odd number of electrons on the island requires an extra energy cost of the superconducting gap, the ground state consists of an even number of electrons. This effect has been directly observed as an even-odd asymmetry in aluminum islands \cite{TuominenExperimentalevidenceparitybased1992, LafargeMeasurementevenoddfreeenergy1993, EilesEvenoddasymmetrysuperconductor1993,  AverinSingleelectronchargingsuperconducting1992a}. On the other hand, in topological superconductors zero energy Majorana bound states exist that can accommodate an unpaired electron without any additional energy. In this case, the degeneracy can occur between states with even and odd number of electrons. This even-odd degeneracy underlies the phenomenon of electron teleportation \cite{FuElectronTeleportationMajorana2010}, which involves phase-coherent transport of a single electron via the spatially separated Majorana modes. A recent experiment \cite{AlbrechtExponentialprotectionzero2016a} on proximitized InAs island connected to two normal leads via tunnel junctions observed a transition from resonant Cooper pair transport to single-electron transport above a critical magnetic field, which is broadly consistent with the scenario of transition from conventional to topological superconducting island \cite{vanHeckConductanceproximitizednanowire2016a, LutchynTransportMajoranaIsland2017}.

The degeneracy between the two charge states of the superconducting islands - $2N$ and $2N+2$ ($2N+1$) in conventional (topological) case - can be a source for Kondo-type phenomena. These degenerate levels can be represented as a pseudospin-1/2 object, which enables observation of phenomena related to the multichannel Kondo effect \cite{NozieresKondoeffectreal1980, CoxExoticKondoeffects1998}. In the topological case, when the superconductor is tuned into charge degeneracy, it has been shown \cite{MichaeliElectronteleportationstatistical2017a, HerviouManyterminalMajoranaisland2016} that the system exhibits quantized DC conductance $G_{ij\mathrm{DC}}=\frac{2e^2}{N h}$ in $T=0$ limit for $N$ Majorana modes coupled to $N$ normal leads by mapping the model onto quantum Brownian motion (QBM) on a honeycomb lattice \cite{YiQuantumBrownianmotion1998, YiResonanttunnelingmultichannel2002}. In the s-wave case, the setup based on a two terminal island at charge degeneracy has recently been shown \cite{PustilnikQuantumCriticalityResonant2017} to map to two channel Kondo problem. These parallel developments not only enable a direct comparison of the properties of both conventional and topological setup, but also provide an attractive new platforms for studies of quantum criticality. Highly tunable nanostructures can serve as a window into the world of strongly-correlated electron systems and so they are intensively studied \cite{IftikharTwochannelKondoeffect2015, IftikharTunablequantumcriticality2018, PotokObservationtwochannelKondo2007, MebrahtuQuantumphasetransition2012, KellerUniversalFermiliquid2015} in order to extract the essence of the physical phenomena without the picture being blurred by the complexity of the real materials.

Motivated by the above results on Majorana and conventional superconducting islands, we expand on these studies by comparing and contrasting the charge Kondo effects due to even-odd and even-even degeneracies in both types of mesoscopic islands using bosonization and numerical renormalization group (NRG) methods. We provide a mapping of the $N$ terminal conventional superconductor island model to $N$ channel charge Kondo problem in the bosonization language and then examine the differences in the treatment of the Majorana island. For non-interacting leads in the topological model the system flows to strong coupling fixed point, as opposed to the flow towards intermediate coupling in the conventional case (see Fig.\ref{fig:flow_diagram}). The non-Fermi liquid fixed point of Majorana island is robust to channel coupling asymmetry (in contrast to anisotropy being a relevant perturbation at the intermediate fixed point in the conventional system). These differences between both types of islands in transport properties are due to the crucial distinction of charge $2e$ transfer in the Andreev processes in the conventional case versus charge $e$ transfer by single electron tunneling in the topological island \cite{FuElectronTeleportationMajorana2010}. In the topological case, while each tunneling process transfers a single electron charge $e$, due to the statistical transmutation \cite{MichaeliElectronteleportationstatistical2017a} the system behaves as if charge-$e$ boson was transferred, which enables a nontrivial mapping to a Kondo model \cite{MichaeliElectronteleportationstatistical2017a, HerviouManyterminalMajoranaisland2016}.

Using numerical renormalization group we first support our bosonization results at $T=0$ by calculating the residual entropy and conductance matrix elements. For the conventional island we confirm that the DC conductance in $T=0$ approaches the predicted value of $2$ and $\frac{8}{3}\sin^2(\frac{\pi}{5})\,e^2/h$ for 2 and 3 terminals, respectively. In the Majorana setup for 3 terminals we obtain the anticipated DC conductance of $\frac{2e^2}{N h} = 2/3\,e^2/h$, which is robust against the tunnel coupling anisotropy (even if all three couplings are different) and moving away from the charge degeneracy point. More importantly, we go beyond the zero temperature limit and obtain the full crossover to non-Fermi liquid fixed points in both cases. In the conventional setup, our results establish the s-wave island as a new platform for studying physics of the intermediate multichannel Kondo fixed point. For Majorana islands, we demonstrate that the transition at the charge degeneracy point happens at a much higher temperature than in the Coulomb valley regime of topological Kondo effect studied previously \cite{BeriTopologicalKondoEffect2012, AltlandMultiterminalCoulombMajoranaJunction2013, GalpinConductancefingerprintMajorana2014,
AffleckTopologicalsuperconductorLuttinger2013, AltlandBetheansatzsolution2014, AltlandMultichannelKondoImpurity2014, ZazunovTransportpropertiesCoulomb2014, ErikssonTunnelingspectroscopyMajoranaKondo2014a, ErikssonNonFermiLiquidManifoldMajorana2014, KashubaTopologicalKondoEffect2015, PikulinLuttingerliquidcontact2016a, MeidanTransportsignaturesinteracting2016, PluggeKondophysicsquasiparticle2016}. Our results facilitate the experimental observation of quantized conductance in future. For the three terminal case, we predict a non-trivial crossover between the regimes dominated by two and three leads with an intermediate DC conductance plateau at 2/3 $e^2/h$, which emerges at sufficiently low temperature while tuning the tunnel coupling of the third lead. This, together with the aforementioned robustness to variation in setup parameters, provides an experimental signature that can be used to verify our claims for the Majorana island.

\section{Models}

\begin{figure}
\includegraphics[width=0.49\textwidth]{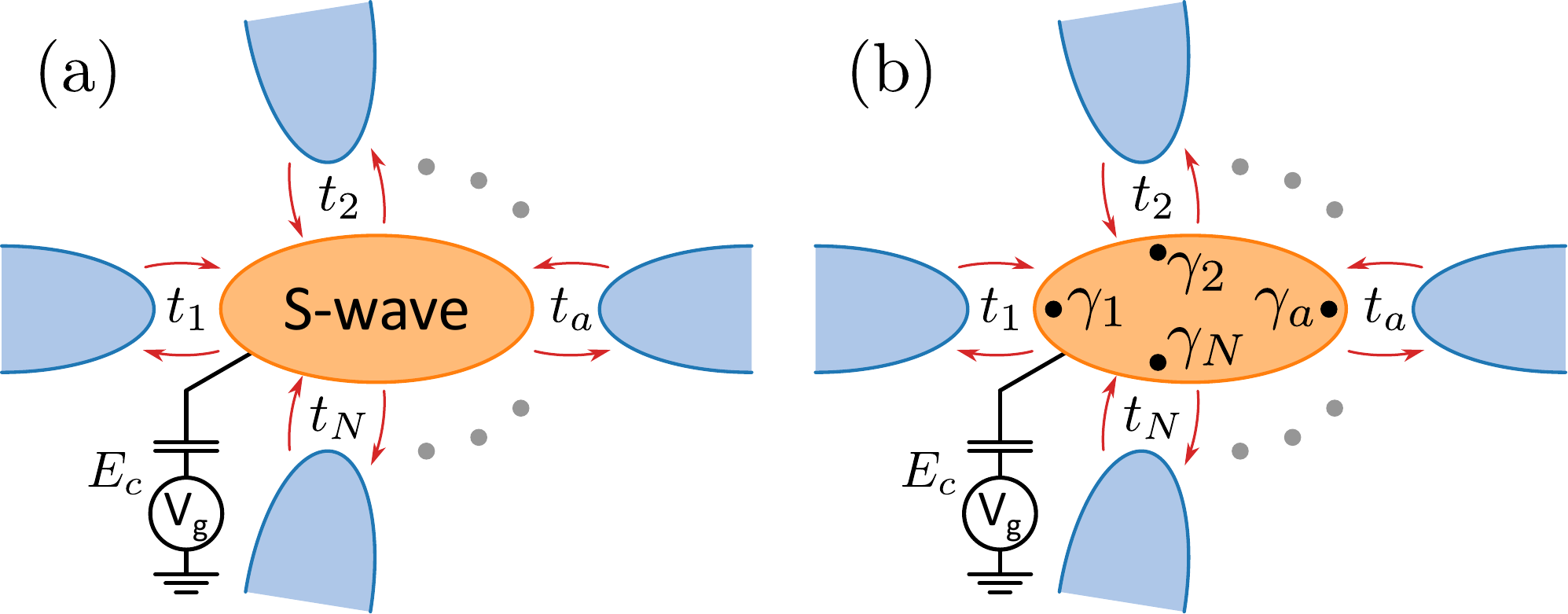}
\caption{\label{fig:setup}(Color online) Multiterminal (a) conventional and (b) Majorana islands at charge degeneracy (charging energy is controlled by a gate). Both islands are connected to $N$ normal leads (blue) via either Andreev reflection or tunneling into Majorana zero modes.}
\end{figure}

In this work, we consider two independent types of setups with multiterminal superconducting islands (Fig.~\ref{fig:setup}). We begin by describing the full Hamiltonian of the systems analyzed in the following sections, which consists of three parts:
\begin{equation}
H = H_C + H_L + H_T
\end{equation}
The central point of both setups considered in this paper is a mesoscopic superconducting island, either of ordinary s-wave or topological nature with a gap $\Delta$ that is the largest energy scale of the problem. In the s-wave case, there are no quasiparticle excitations inside of the superconducting gap and so in the usual BCS formalism introducing an odd number of electrons into the island requires an energy cost of $\Delta$. On the other hand, the topological superconductor hosts an even number of zero energy Majorana bound states and this allows to put an additional electron into the island without paying the extra energy. Since we are studying a mesoscopic superconducting grain that is not grounded, we also have to consider the charging effects which arise due to Coulomb interactions. The electrostatics can be taken into account by including into the Hamiltonian a term:
\begin{equation}
\label{eq:charging_energy}
H_C = E_C (\hat{N} - N_g)^2
\end{equation}
where $E_C$ is the charging energy related to the capacitance of the island, $\hat{N}$ is the number of charges in the superconductor and $N_g$ is the potential determined by the external gate. This tunability gives rise to a possible degeneracy between the two charge states of the island. However, the number of charges in the degenerate states differs in both considered cases. For the ordinary superconductor, since we are working in the regime where $E_C \ll \Delta$, we can consider the states with odd number of electrons to be unfavorable energetically and so when we set $N_g$ to be an odd integer $2N_0+1$, according to \eqref{eq:charging_energy}, the states with $2N_0$ and $2N_0 + 2$ electrons will be degenerate and lowest in energy. The situation is different in the topological superconductor, where there is no additional energy cost for the states with an odd number of electrons. There we can set $N_g$ to $2N_0+\frac{1}{2}$ and then states with $2N_0$ and $2N_0+1$ are degenerate. At very low temperatures we can then restrict our Hilbert space to just those pairs of charge states of the island. The subspace of charge states can then be described by a spin-$1/2$ object, with $s_z$ eigenstates corresponding to $2N_0/2N_0+1$ or $2N_0/2N_0+2$ states. Then the slight deviation from charge degeneracy point can be taken into account by introducing Zeeman-like term $\delta s_z$ into Hamiltonian, where $\delta$ can be tuned by the external gate.

A common part of both setups is a set of $N$ normal leads, which are tunnel coupled to the superconductor. In the ordinary superconductor setup they are described by the Hamiltonian of spinful fermions with dispersion linearized close to the Fermi energy:

\begin{equation}
H_L = - i v_F\sum_{a,\sigma=\uparrow/\downarrow}^N \int_0^{\infty} dx \psi^\dagger_{a,R,\sigma} \partial_x \psi_{a,R,\sigma} - \psi^\dagger_{a,L,\sigma} \partial_x \psi_{a,L,\sigma}
\end{equation}
where $\psi_{a,r=L/R,\sigma=\uparrow/\downarrow}(x)$ are operators annihilating left/right moving modes with spin $\sigma$ at the point $x$ of the lead $a$, combining into $\psi_{a,\sigma}(x) = \psi_{a,R,\sigma}(x) e^{i k_F x} + \psi_{a,L,\sigma} e^{-i k_F x}$. However, a difference arises in the topological case, because Majorana states couple only to one of two spin components \cite{LawMajoranaFermionInduced2009a, HaimSignaturesMajoranaZero2015}. This allows to drop the spin index in this case and consider spinless fermions.

The leads are semi-infinite, ending at $x=0$ where they are in contact with the superconductor. The exact form of the tunneling Hamiltonian depends then on the type of superconductor. In the case of the s-wave superconductor the charge transfer into the island will occur due to the Andreev processes in which incident electron in the lead is reflected as a hole and at the same time a single Cooper pair is added to the superconductor. Using the spin-$1/2$ representation of the charge state of the island we can write the tunneling Hamiltonian as:
\begin{equation}
\label{eq:tun_ham_swave}
H_T = \sum_{a=1}^N t_a (\psi_{a,\uparrow}^\dagger(0) \psi_{a,\downarrow}^\dagger(0) s^- + \psi_{a,\downarrow}(0) \psi_{a,\uparrow}(0) s^+)
\end{equation}
where we are either adding or removing two electrons of opposite spin at $x=0$ point of the lead $a$ and at the same time changing the charge state of the island between $2N_0$ and $2N_0 + 2$. In writing this Hamiltonian we assumed that the superconducting island is large enough so that the crossed Andreev reflection is suppressed. On the other hand, in the case of topological superconductor, the tunneling will occur into the Majorana zero modes. We also use spin-$1/2$ representation of the charge state, with transitions between $2N_0$ and $2N_0+1$ states. Then the tunneling Hamiltonian has the form \cite{MichaeliElectronteleportationstatistical2017a, HerviouManyterminalMajoranaisland2016}

\begin{equation}
\label{eq:tun_ham_majorana}
H_T = \sum_{a=1}^N (t_a \psi_a^\dag s^- \gamma_a + H.c.)
\end{equation}
where $t_a$ are tunnel couplings to the leads, $\psi_a^\dag$ are creation operators at the end of the leads and $\gamma_a = \gamma_a^\dag$ are Majorana operators.

\section{Bosonization analysis}

Both setups can now be studied using bosonization by transforming the normal leads into Luttinger liquids, spinful in the case of s-wave island and spinless when leads are coupled to Majorana zero modes. We derive the results for the ordinary superconductor and then highlight the differences that arise in the Majorana setup \cite{MichaeliElectronteleportationstatistical2017a, HerviouManyterminalMajoranaisland2016}.
\subsection{S-wave island}

After spinful bosonization, the Hamiltonian of the leads has now the form \cite{QuantumPhysicsOne2004}:
\begin{equation}
H_L = \sum_{a=1}^N \frac{v}{2 \pi} \sum_{j=\rho,\sigma} \int_{0}^\infty dx K_j (\nabla \theta_{a,j})^2 + \frac{1}{K_j} (\nabla \phi_{a,j})^2
\end{equation}
where we have used the following convention:
\begin{equation}
\label{eq:bosonization_identity_spinful}
\psi_{a, r,\sigma}(x) = \frac{U_{a, r, \sigma}}{\sqrt{2 \pi \alpha}}  e^{-\frac{i}{\sqrt{2}} (r \phi_{a,\rho}(x) - \theta_{a,\rho}(x) + \sigma(r \phi_{a,\sigma}(x) - \theta_{a,\sigma}(x)))}
\end{equation}
with $\alpha$ being short distance cut-off and $U_{a, r,\sigma}$ are the Klein factors. Using \eqref{eq:bosonization_identity_spinful} we can now express the tunneling Hamiltonian using bosonic fields. Since the lead ends at $x=0$, we can impose the boundary condition $\psi_{a,L,\sigma}(0) = \psi_{a,R,\sigma}(0)$. This in turn means that $\phi_{\rho/\sigma}(0) = 0$ and that Klein factors for right and left movers of each spin are equal: $U_{a, R,\sigma}=U_{a, L,\sigma} = U_{a, \sigma}$. Combining all of this together we express the tunneling Hamiltonian \eqref{eq:tun_ham_swave} as:
\begin{equation}
H_T = \sum_{a=1}^N \frac{2 t_a}{\pi \alpha} (U_{a,\uparrow} U_{a, \downarrow}e^{- i \sqrt{2} \theta_{a,\rho}(0)} s^- + H.c.)
\end{equation}
We can form a parity operator from the Klein factors $p_a = i U_{a,\uparrow} U_{a, \downarrow}$ and since $p_a^2 = 1$ we can use the identity $e^{i \gamma p_a} = \cos(\gamma) + i p_a \sin(\gamma)$. For $\gamma=\frac{\pi}{2}$ this translates to $i p_a = e^{i \frac{\pi}{2} p_a}$, so we have $U_{a,\uparrow} U_{a, \downarrow} = - i p_a = e^{-i \frac{\pi}{2} p_a}$. Thus the Klein factors translate to a phase shift, which can be absorbed into the bosonic field, because the parity in each lead is fixed as the only allowed tunneling process transfers pairs of electrons. The final form of the tunneling Hamiltonian is then:
\begin{equation}
H_T = \sum_{a=1}^N \frac{2 t_a}{\pi \alpha} (e^{- i \sqrt{2} \theta_{a,\rho}(0)} s^- + e^{ i \sqrt{2} \theta_{a,\rho}(0)} s^+)
\end{equation}

Therefore, both bosonic fields from the spin sector ($\theta_\sigma$ and $\phi_\sigma$) and $\phi_\rho$ are not present in the tunneling Hamiltonian and are present only in the quadratic part of the action. This means that we can integrate them out from the imaginary time action. Moreover, the field $\theta_\rho$ is taken only at $x=0$ in $H_T$, so we can also integrate it out away from $x=0$. After this procedure, we obtain the imaginary time action:

\begin{align}
S^\mathrm{s-wave} &= S_0^\mathrm{s-wave} + S_T^\mathrm{s-wave} \\
S_0^\mathrm{s-wave} &= \sum_{a=1}^N \frac{K_\rho}{2 \pi} \int \frac{d\omega}{2 \pi} |\omega| |\theta_{a,\rho}(\omega)|^2 \\
\label{eq:S_T_swave}
S_T^\mathrm{s-wave} &= \sum_{a=1}^N \int_0^\beta d\tau  \frac{J_{\perp,a}}{2} (e^{- i \sqrt{2} \theta_{a,\rho}(0)} s^- + H.c)
\end{align}

In anticipation of the connection of this action to the multichannel Kondo problem we introduced the notation for the coupling $ J_{\perp,a} = \frac{4 t_a}{\pi \alpha}$.

\subsection{Majorana island}

The procedure of obtaining the effective boundary action in the case of Majorana island is essentially the same, with the important distinction that the leads now contain effectively spinless electrons and so the bosonization identity now takes form:

\begin{equation}
\label{eq:bosonization_identity_spinless}
\psi_{a, r}(x) = \frac{U_{a, r}}{\sqrt{2 \pi \alpha}}  e^{-i (r \phi_{a}(x) - \theta_{a}(x))}
\end{equation}
with $\alpha$ again being the short distance cut-off and $U_{a, r}$ are the Klein factors. The bosonized Hamiltonian of the leads is:

\begin{equation}
H_L = \sum_{a=1}^N \frac{v}{2 \pi} \int_{0}^\infty dx K (\nabla \theta_{a})^2 + \frac{1}{K} (\nabla \phi_{a})^2
\end{equation}

and tunneling Hamiltonian is:

\begin{equation}
H_T = \sum_{a=1}^N \frac{2 t_a}{\sqrt{2\pi \alpha}} (e^{- i \theta_{a}(0)} s^- + e^{ i \theta_{a}(0)} s^+)
\end{equation}
where the Klein factors hybridized with Majorana operators in a process of statistical transmutation \cite{BeriMajoranaKleinHybridizationTopological2013, MichaeliElectronteleportationstatistical2017a}.

When the bosonic field is integrated out away from $x=0$, we obtain the imaginary time action:

\begin{align}
S^\mathrm{M} &= S_0^\mathrm{M} + S_T^\mathrm{M} \\
S_0^\mathrm{M} &= \sum_{a=1}^N \frac{K}{2 \pi} \int \frac{d\omega}{2 \pi} |\omega| |\theta_{a}(\omega)|^2 \\
S_T^\mathrm{M} &= \sum_{a=1}^N \int_0^\beta d\tau \frac{J_{\perp,a}}{2} (e^{- i \theta_{a}(0)} s^- + e^{ i \theta_{a}(0)} s^+) \label{eq:S_T_topo}
\end{align}
This time we made the identification $ J_{\perp,a} = \frac{4 t_a}{\sqrt{2\pi \alpha}}$. It is interesting to make a comparison between tunneling parts of the action for both cases. Eqs. \eqref{eq:S_T_swave} and \eqref{eq:S_T_topo} have virtually the same form, apart from the factor of $\sqrt{2}$ in the exponent for the s-wave superconducting island model. One can then perform the following transformation of the action \eqref{eq:S_T_topo}: $\theta_a \rightarrow \sqrt{2} \tilde{\theta}_a$. In order to keep the quadratic part of the action the same under this transformation, we also have to rescale the Luttinger parameter $K$: $K \rightarrow \tilde{K}/2$. This means that the topological system will behave exactly the same as the ordinary one, but with interaction parameter rescaled by factor of 2. This bears important consequences for the flow diagram of perturbative RG close to the non-interacting value of $K=1$.

\subsection{Perturbative renormalization group analysis}

\begin{figure}
\includegraphics[width=0.49\textwidth]{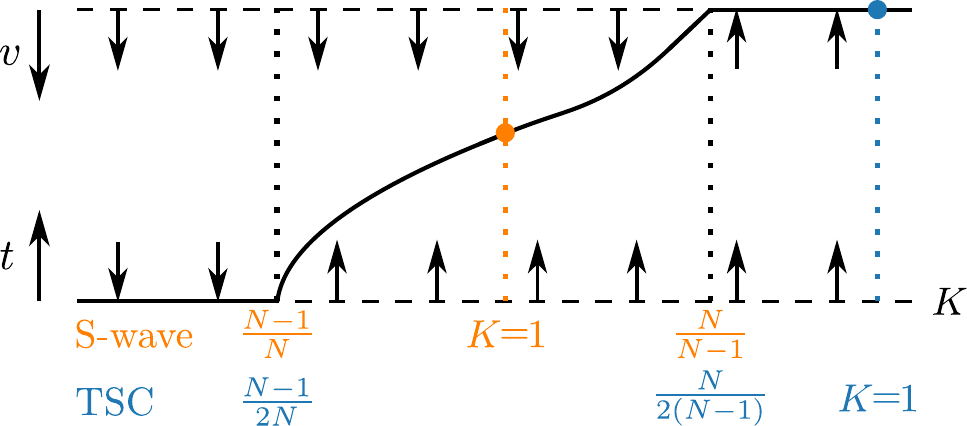}
\caption{\label{fig:flow_diagram}(Color online) Renormalization group flow for s-wave and topological superconducting (TSC) islands conncted to $N<5$ leads as a function of the Luttinger parameter $K$. Stable and unstable fixed points are depicted as solid and dashed lines, respectively. The bottom line corresponds to the limit of weak tunneling ($t\rightarrow 0$) and the top line corresponds to weak periodic potential ($v_{\vec{G}} \rightarrow 0$). Due to rescaling of $K$, in the Majorana island case for non-interacting leads ($K=1$) is of strong coupling nature, compared to intermediate coupling for s-wave island.}
\end{figure}

Because we have shown that there exists a direct correspondence between the actions for both s-wave and topological islands, it is sufficient to perform perturbative renormalization group analysis of the action of the s-wave setup and then recover the behavior of the Majorana island by substituting $\tilde{K}$ for $K_\rho$. During the RG procedure an additional term is generated which is proportional to $\partial_x \phi_{a,\rho}$, even if its coupling is initially zero. Therefore we add it into the action right from the beginning with $J_z$ coupling:
\begin{equation}
S_z = - \sum_{a=1}^N \frac{v}{\sqrt{2}} J_{z} s_z \partial_x \phi_{a,\rho}
\end{equation}

With this additional term and the relabeling of the couplings done in the previous section, the complete action for our problem has exactly the same form as the anisotropic multichannel Kondo problem action \cite{YiQuantumBrownianmotion1998, YiResonanttunnelingmultichannel2002}. Therefore the analysis steps follow directly from the standard procedure applied previously to the Kondo problem.

We begin by considering the isotropic limit when all $J_{\perp,a} = J_\perp$. In such a case the RG equations for the couplings are:

\begin{equation}
\frac{dJ_z}{dl} = J_\perp^2(\frac{1}{K_\rho} - \frac{N}{2} J_z)
\end{equation}
\begin{equation}
\frac{dJ_\perp}{dl} = (1-\frac{1}{K_\rho})J_\perp + J_z J_\perp(1-\frac{N K_\rho}{4} J_z)
\end{equation}
Those equations are exact in $J_z$ and perturbative in $J_\perp$. We notice that in the isotropic case the couplings flow to the Toulouse fixed point, where $J_z$ becomes $\frac{2}{N K_\rho}$ and its flow stops. This means that we can perform a unitary transformation and eliminate the $\partial_x \phi_\rho$ term from the Hamiltonian:

\begin{equation}
U = e^{i K_\rho J_z \sqrt{\frac{N}{2}} \Theta(0) s_z}
\end{equation}

\begin{equation}
U^\dagger H U = H_L - \sum_{a=1}^n \frac{J_\perp}{2} (e^{- i \sqrt{2} (\theta_{a,\rho}(0)- \frac{1}{\sqrt{N}} \Theta(0))} s^- + H.c.)
\end{equation}
where $\Theta(0)= \frac{1}{\sqrt{N}} \sum_j \theta_{j,\rho}(0)$ is the global mode.

We can now determine the fixed points of the problem and understand them using the quantum Brownian motion correspondence. In QBM picture, the state of the system is described as a position of a fictitious particle placed in periodic potential with dissipative environment. This enables approaching the problem from two dual perspectives: tunneling between the minima of a strong periodic potential and free motion with weak potential as a perturbation. To make the mapping clearer we can write the tunneling operators in the action as $e^{- i \sqrt{2} \vec{\theta}_{\rho} \vec{R}_0^{(a)}} s^-$, where $\vec{\theta}_\rho = (\theta_{1,\rho}, ..., \theta_{N,\rho})$ and $\vec{R}_0^{(a)}$ is a vector with 1 on the $a$th component and 0 on the rest. In this notation one can think of $\vec{\theta}_\rho$ as the momentum of the particle and the number of charges in the leads (which is a variable conjugate to $\vec{\theta}_\rho$) describes the position of the particle. When the periodic potential is strong, the particle is mostly localized in the minima of the potential which are connected by the lattice vectors $\vec{R}_0^{(a)}$ and only occasionally tunnels between. Since we have charging energy in our setup and the island can only accommodate a single additional Cooper pair, the total number of charges in the leads $N_\mathrm{tot}$ can only change between $N / N+2$ and the particle's motion is restricted to two planes in the $N$ dimensional space. The corresponding lattices are 1D zig-zag chain for $N=2$ channels and $N-1$ dimensional hyperhexagonal lattice for $N>2$. Both lattice types are non-symmorphic with two atom basis, which corresponds to presence or absence of the additional Cooper pair in the superconducting island.

In this language, the global mode introduced after the unitary transformation at the Toulouse fixed point corresponds to the product of $\vec{\theta}_\rho$ and the vector $\vec{R}_\perp = \frac{1}{\sqrt{N}} (1, ..., 1)$ perpendicular to the planes to which the particle motion is confined. The tunneling operators after the transformation are $e^{- i \sqrt{2} \vec{\theta}_{\rho} \vec{R}_\parallel^{(a)}} s^-$, with $\vec{R}_\parallel^{(a)} = \vec{R}_0^{(a)} - \frac{1}{\sqrt{N}}\vec{R}_\perp$. The scaling dimension of the tunneling operator is then:
\begin{equation}
\Delta[e^{- i \sqrt{2} \vec{\theta}_{\rho} \vec{R}_\parallel^{(a)}} s^-] = \frac{|\vec{R}_\parallel^{(a)}|^2}{K_\rho} = \frac{1}{K_\rho}( 1-\frac{1}{N})
\end{equation}
Therefore determining whether tunneling operator is relevant depends on the Luttinger parameter $K_\rho$ - the condition for relevancy is $K_\rho>\frac{N-1}{N}$. Importantly, this means that for non-interacting electrons ($K_\rho=1$), for all $N$ the tunneling operator is relevant and the system will be flowing in direction of decreasing periodic potential strength, away from the localized fixed point. Since the coupling increases substantially, the perturbation theory breaks down and we need to find the stable fixed point properties in another way. To do this, we can use the dual perspective of looking at the QBM as a free motion with weak potential perturbation. In this case, the periodic potential can be expressed using its Fourier components $V(\vec{r}) = \sum_{\vec{G}} v_{\vec{G}} e^{i \vec{G} \vec{r}}$, where $\vec{G}$ are reciprocal lattice vectors of the honeycomb lattice. Then the scaling dimension of the most relevant $v_{\vec{G}}$ (corresponding to the shortest $\vec{G}$) is given by \cite{MichaeliElectronteleportationstatistical2017a}:
\begin{equation}
\Delta [e^{i \vec{G} \vec{r}}] = K_\rho |\vec{G}|^2 = K_\rho \left(1-\frac{1}{N} \right)
\end{equation}
Again, relevancy of the periodic potential perturbation depends on the value of $K_\rho$. The criterion in this case is $K_\rho<\frac{N}{N-1}$, which for non-interacting leads is always satisfied: the periodic potential is a relevant perturbation to the free motion fixed point. Therefore, there have to be additional fixed points between the localized and free motion, including at least one stable. This analysis is summarized for $N<5$ in Fig.~\ref{fig:flow_diagram}, which indicates stable and unstable fixed points as solid and dashed lines, respectively. For $N\geq5$ there exists another unstable intermediate coupling fixed point that has been analyzed in more detail by Yi \cite{YiResonanttunnelingmultichannel2002}. The stable intermediate coupling fixed point for non-interacting leads has been studied using conformal field theory in the context of multichannel Kondo problem \cite{AffleckCriticaltheoryoverscreened1991, AffleckExactconformalfieldtheoryresults1993}. Applying those results to our model we can immediately find the zero temperature residual entropy:
\begin{equation}
S_\mathrm{imp}(T=0) = \ln\left(2 \cos\left(\frac{\pi}{N+2} \right)\right)
\end{equation}
Moreover, we can also deduce the zero temperature conductance matrix elements to be:
\begin{equation}
\label{eq:SC_zero_temp_cond}
G_{ij}(T=0) = 8 \sin^2 \left( \frac{\pi}{N+2} \right) \left(\frac{1}{N} - \delta_{ij} \right)\frac{e^2}{h}
\end{equation}
The important distinguishing feature is that compared to the Kondo problem, conductance matrix element here is quadrupled. Each Andreev reflection process transfers the charge of $2e$ between the leads and superconducting island, corresponding to doubling of the current compared to conventional charge Kondo effect. This current operator is then used in the Kubo formula to obtain conductance as current-current correlation function and the doubling translates in this way to quadrupling of $G_{ij}(T=0)$. The conformal field theory gives also the scaling dimension of leading irrelevant operator at the intermediate fixed point that translates to the leading temperature correction to $G_{ij}(T=0)$:
\begin{equation}
\label{eq:temp_corr_cond}
G_{ij}(T) = G_{ij}(T=0) \left(1 - c \left(\frac{T}{T_K}\right)^{\lambda}\right)
\end{equation}
where $\lambda$ is 1 for $N=2$ \cite{MitchellUniversalityScalingCharge2016a, LandauChargeFractionalizationTwoChannel2018, FurusakiTheorystronginelastic1995a} and $\frac{2}{5}$ for $N=3$ \cite{AffleckExactconformalfieldtheoryresults1993, IftikharTunablequantumcriticality2018, CoxExoticKondoeffects1998}, $c$ is a constant on the order of unity, $T_K$ is the Kondo temperature. One important characteristic of the intermediate fixed point is that it is unstable to channel coupling asymmetry \cite{AffleckRelevanceanisotropymultichannel1992, FabrizioCrossoverNonFermiLiquidFermiLiquid1995a}: when one of the couplings is increased, the system will flow to Fermi liquid fixed point that describes the single channel Kondo model and when one of the couplings is decreased, the system will behave as $N-1$ channel setup in low temperatures. In general, the asymmetric system will behave as $N_\mathrm{max}$ channel setup at low energy scales, where $N_\mathrm{max}$ is the number of leads with the largest coupling value. This constitutes a significant difficulty in performing experiments that verify the theoretical claims listed above.

Now we can turn to the case of Majorana island in which we have to substitute $K_\rho \rightarrow \tilde{K} = 2 K$. This change essentially shifts the flow diagram and redefines the condition for relevancy of tunneling and weak periodic potential operators, which are now $K>\frac{N-1}{2N}$ and $K<\frac{N}{2(N-1)}$, respectively. This is also indicated in Fig.~\ref{fig:flow_diagram} (which again is valid for $N<5$ with a new unstable fixed point appearing for $N\geq 5$). The redefinition of relevancy condition brings about a crucial change for the non-interacting leads: while the tunneling operator is still relevant for $K=1$, the weak periodic potential becomes irrelevant for all $N$. This means that the free motion fixed point becomes stable and that conductance will assume maximum value allowed by the charge conservation. Remembering that in Majorana island the tunneling processes carry charge of $1e$, we find that the conductance is:
\begin{equation}
G_{ij}(T=0) = 2 \left(\frac{1}{N} - \delta_{ij} \right)\frac{e^2}{h}
\end{equation}

The weak periodic potential becomes now the leading irrelevant operator and its scaling dimension will now determine the exponent of the temperature correction of the conductance:
\begin{equation}
\Delta_\mathrm{irr} = 2\left(1-\frac{1}{N} \right)
\end{equation}
The form of the correction is still described by \eqref{eq:temp_corr_cond}. The change of the nature of the low temperature fixed point comes with another major difference: the channel coupling anisotropy, which corresponds to deformation of the periodic potential becomes an irrelevant perturbation and doesn't cause the system to flow to the Fermi liquid fixed point. This will be explored in more detail in the numerical section.

\section{Numerical results}

To verify the analytical results and obtain a fuller understanding of the crossover regime between the fixed points of the studied models, we employ numerical renormalization group (NRG), a powerful non-perturbative method for obtaining thermodynamics and correlation functions of quantum impurity systems, connected to non-interacting leads \cite{BullaNumericalrenormalizationgroup2008}. As we want to capture the universal physics of this setup, we simplify the problem by assuming that all the leads are identical with bandwidth $2D$ and a flat density of states $\rho = 1/2D$. Then we express the Hamiltonians in the form suitable for calculations as explained in the sections below. We perform the spectral function calculations in the framework of full density matrix NRG \cite{WeichselbaumSumRuleConservingSpectral2007} using complete basis set \cite{AndersRealTimeDynamicsQuantumImpurity2005} in order to properly account for finite temperature effects. We also perform sliding parameter averaging \cite{OliveiraGeneralizednumericalrenormalizationgroup1994} \cite{ZitkoEnergyresolutiondiscretization2009} over 4 values of sliding parameter $z$ to remove spurious oscillations in the results. In the calculations of the s-wave island model we use the discretization parameter $\Lambda=5$ for 2 channel system and $\Lambda=10$ for 3 channel system. To make the 3 channel case numerically tractable, we use the interleaved Wilson chain scheme \cite{MitchellGeneralizedWilsonchain2014a, StadlerInterleavednumericalrenormalization2016}, which requires some fine-tuning of the individual tunnel couplings to obtain the critical behavior. The cut-off energy has been set to $E_\mathrm{cut}=14$ for two leads and $E_\mathrm{cut}=8.5$ for three lead case, with the maximum number of kept states 15 000 and 40 000, respectively. In the Majorana island part, we used $\Lambda=3$ and kept up to 5000 states in each iteration.

To directly relate our results to the experiment, we focus on the DC conductance in our calculations. We work in the framework of the linear response theory and compute AC conductance using the Kubo formula as the correlation function of number of electrons in one lead and current in the other lead. This allows to avoid computation of the delicate limit present in the usual current-current correlation approach (see Appendix A). Finally, we obtain DC conductance as the limit $G_{jk\mathrm{DC}}(T) = \lim_{\omega \rightarrow 0}G_{jk\mathrm{DC}}(\omega, T)$ of the AC conductance.

\subsection{Superconducting island}

\begin{figure*}
\includegraphics[width=0.99\textwidth]{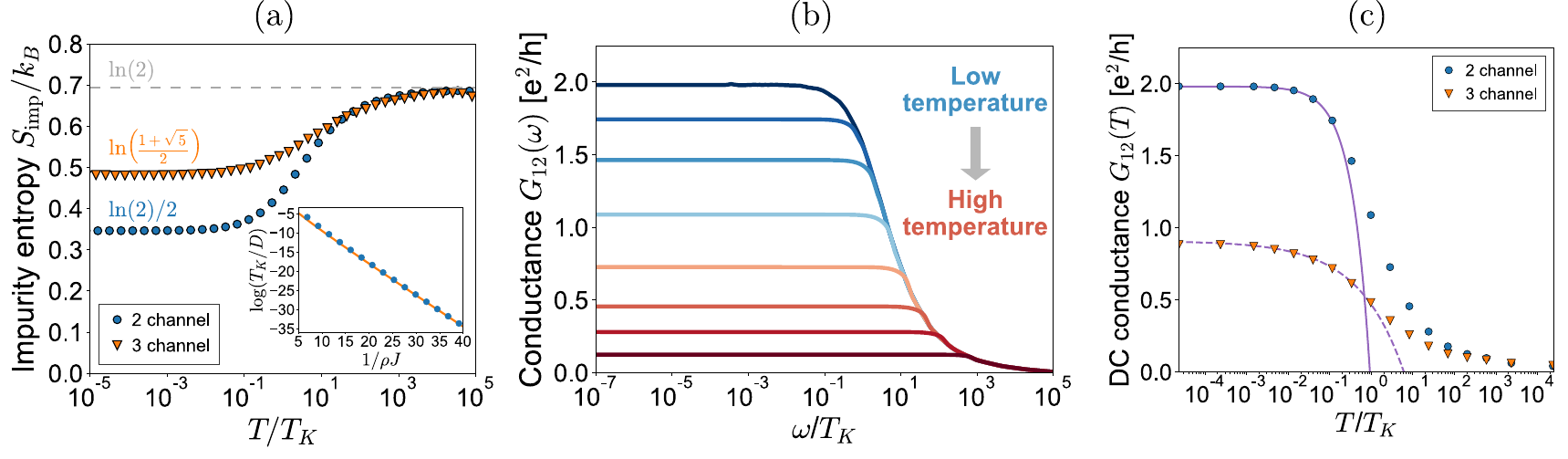}
\caption{\label{fig:SC_NRG}(Color online) (a) The impurity entropy $S_\mathrm{imp}(T)$ curves showing the crossover between local moment ($S_\mathrm{imp}(T)=\ln(2)$) and non-Fermi liquid fixed points ($S_\mathrm{imp}(T)=\ln(\sqrt{2})$ for 2 channels and $\ln(\frac{1+\sqrt{5}}{2})$ for 3 channels). The inset shows that $T_K$ obtained at the center of the crossover (blue points) is given by $T_K \approx \rho J_\perp \exp(\frac{\pi}{4 \rho J_\perp})$ (orange line). (b) $G_{12}(\omega)$ conductance matrix element for two channels in the isotropic case $J_1=J_2=0.15D$ for several temperatures. (c) DC conductance $G_{12\mathrm{DC}}(T)$  with a power-law correction given by the leading irrelevant operator for 2 ($\Delta G\sim T$) and 3 ($\Delta G \sim T^{2/5}$) channels.}
\end{figure*}

We begin by analyzing the numerical results obtained in the case of s-wave superconductor island. Since this setup maps exactly to multichannel Kondo problem (as shown above), which has been studied extensively using NRG, we only highlight that the Andreev reflection Hamiltonian indeed reproduces the key results of Kondo effect. The Hamiltonian used in NRG simulations is:

\begin{equation}
H_\mathrm{SC}^\mathrm{NRG} = H_\mathrm{leads} + \sum_{a=1}^N t_a c^\dagger_{a0\uparrow} c^\dagger_{a0\downarrow} f_{\downarrow} f_{\uparrow} + \mathrm{H.c.}
\end{equation}
where $c_{a0\sigma}$ are the fermionic operators at the end of the Wilson chain that are connected to the superconducting island and $f_\sigma$ describes the pair of fermionic states in the island that are only both occupied or both empty at the same time, simulating the two possible charge states of the island. First, we look at the entropy of the island at low temperatures (Fig.~\ref{fig:SC_NRG}(a)). For both two and three channel cases tuned to the critical point, we observe residual entropy as in the usual Kondo effect. In the two channel case, the entropy flows to $S_\mathrm{2ch}(T=0)=\ln(2)/2$, which is explained by the observation of Emery and Kivelson \cite{EmeryMappingtwochannelKondo1992} that 2 channel Kondo model maps to a resonant level system with only half of the impurity degrees of freedom coupled to the conduction electrons. For three channel case, the entropy flows to $S_\mathrm{3ch}(T=0)=\ln(\frac{1+\sqrt{5}}{2})$, which is consistent with the conformal field theory result and previous numerical studies of regular Kondo effect \cite{MitchellGeneralizedWilsonchain2014a}. The inset of Fig.~\ref{fig:SC_NRG}(a) shows scaling of the Kondo temperature for 2 channel model as the tunnel couplings are varied and this dependence also exactly follows the behavior of the charge Kondo problem \cite{MitchellUniversalityScalingCharge2016a}:
\begin{equation}
T_K/D \sim \rho J \exp(-\frac{\pi}{4 \rho J})
\end{equation}

Next we move on to linear conductance between the normal leads. In Fig.~\ref{fig:SC_NRG}(b) we show the AC conductance matrix element $G_{12}(\omega)$ for several temperatures for the case of 2 channels. All the curves follow the same universal behavior before saturating at their respective DC limit, which in the limit of $T=0$ is equal to 2$e^2/h$ as predicted by the low energy fixed point in the perturbative renormalization group scheme and obtained previously by Pustilnik et al. \cite{PustilnikQuantumCriticalityResonant2017} The values of $G_{12}(\omega\rightarrow 0)$ are then determined for all the remaining temperatures and plotted in Fig.~\ref{fig:SC_NRG}(c), together with corresponding values for the three channel setup. For the three channel setup, the predicted value of $\frac{8}{3}\sin^2(\frac{\pi}{5})\approx 0.92$ is also observed. This calculated temperature dependence is then fitted with the low temperature correction determined by the scaling dimension of leading irrelevant operator at the intermediate fixed point. For $T\ll T_K$ we observe excellent agreement of calculated curve with predicted exponent $\Delta G \sim T$ in the case of two leads and $\Delta G \sim T^{2/5}$ in the case of three leads.

All of the results described above are unstable with respect to tunnel coupling anisotropy, so if the values of $t_a$ are detuned from a common value, the system in general flows to the Fermi liquid fixed point of the single channel Kondo model as expected \cite{AffleckRelevanceanisotropymultichannel1992}.

\subsection{Majorana island}

In the numerical analysis of the Majorana island model we limit our considerations to the first nontrivial case with $N=3$ leads.  In such a system, the dimension of the Hilbert space is 4 corresponding to 4 Majorana modes. It is then divided into 2 two-dimensional subspaces labeled by fermion parity of the island. To transform the Hamiltonian to a form suitable for NRG calculations, we introduce a spinless fermion $f^\dagger$ on the island to distinguish the two subspaces. Each of the two dimensional parity subspaces is then described by a spin-1/2 impurity $\vec{\sigma}$. We note that this spin-1/2 object is a different one than $\vec{s}$ used in the bosonization treatment, which was related to different charge states of the island. Then Hamiltonian has the following form

\begin{equation}
\label{eq:nrg_hamiltonian}
H = H_{\mathrm{leads}} + \sum_{j=1}^3 (t_j \psi_j^\dag \sigma_j f + H.c.) + \delta (f^\dag f - \frac{1}{2}).
\end{equation}

We then define the level broadening $\Gamma = \rho t_\mathrm{avg}^2$, where $t_\mathrm{avg}$ is the average tunnel coupling between the island and the leads. Even though Majorana hybridization is a relevant perturbation in our model, in most of our calculations we neglect it, because the recent experiments show that minimizing the hybridization by using sufficiently long nanowires is possible and allows for performing satisfactory measurements. However, in order to test this assumption we performed some calculations with an additional term $H_\mathrm{hyb} = b_{jk} i \gamma_j \gamma_k$. In our mapping of the Majoranas to a spin-1/2 object, this translates to $H_\mathrm{hyb} = \vec{K}\cdot\vec{\sigma}$, an effective magnetic field for this spin. Hamiltonian of Eq. \ref{eq:nrg_hamiltonian} is now suitable for NRG treatment.

\begin{figure*}
\includegraphics[width=0.75\textwidth]{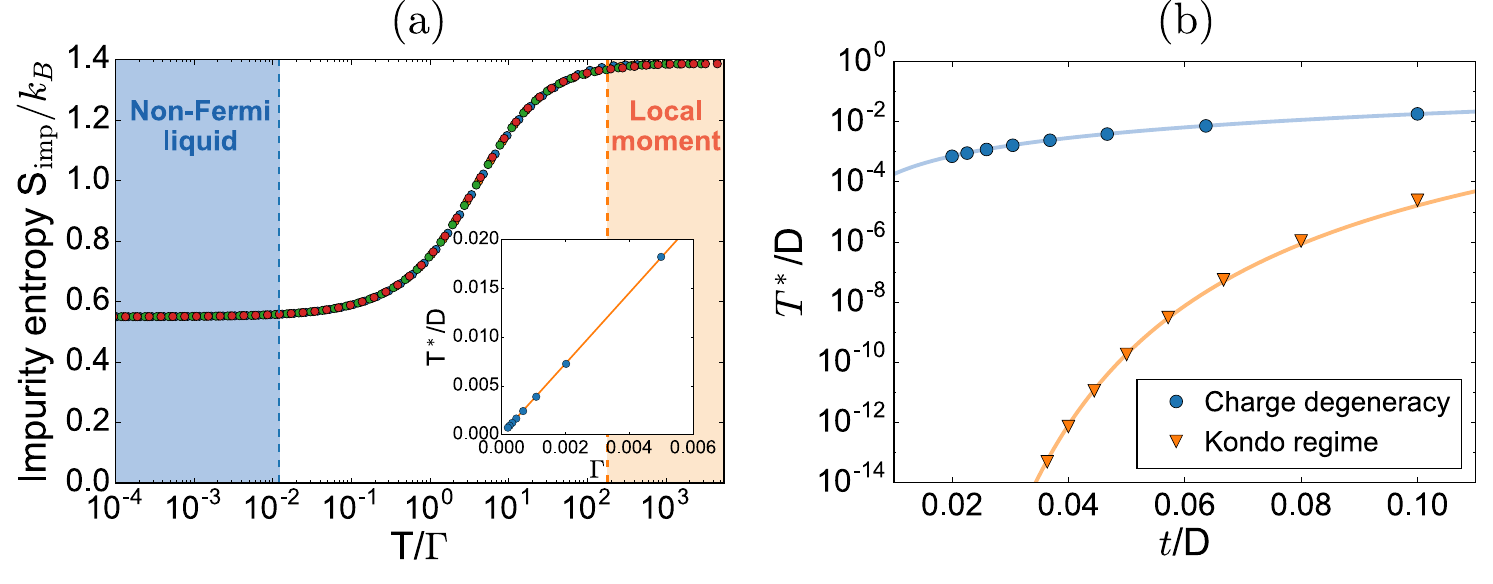}
\caption{\label{fig:entropy}(Color online) (a) The collapsed (for $T$ in units of $\Gamma$) island's entropy $S_\mathrm{imp}(T)$ curves showing the crossover between local moment ($S_\mathrm{imp}(T)=\ln(4)$) and non-Fermi liquid fixed points ($S_\mathrm{imp}(T)=\ln(\sqrt{3})$). The inset shows the linear relation $T^* = c \Gamma$ with $c \approx 3.60$ obtained from fitting. (b) Crossover temperature comparison between charge degeneracy point and the topological Kondo regime for several values of $t=t_1=t_2=t_3$. The temperature at charge degeneracy point is at least 3 orders of magnitude higher than in the Kondo regime.}
\end{figure*}

\begin{figure*}
\includegraphics[width=0.99\textwidth]{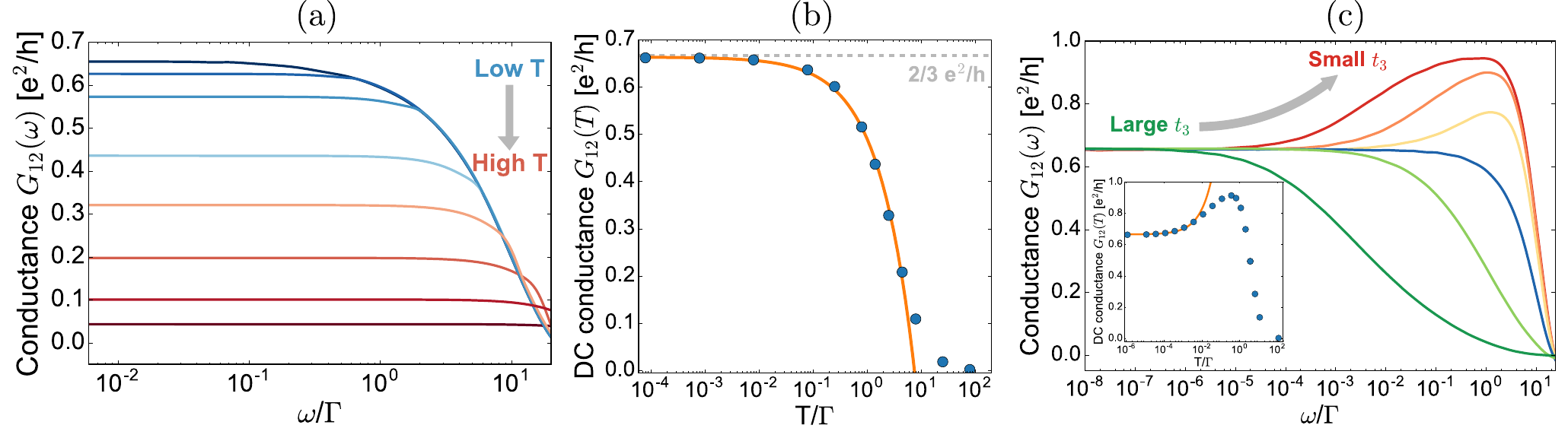}
\caption{\label{fig:anisotropic}(Color online) (a) $G_{12}(\omega)$ conductance matrix element in isotropic case $t_1=t_2=t_3=0.1D$ for several different temperatures showing a universal behavior. c) $G_{12\mathrm{DC}}(T)$ DC conductance with a fit of the universal temperature power-law correction with exponent of 2/3. (c) $G_{12}(\omega)$ conductance matrix element in fully anisotropic case $t_1=0.0475D$, $t_2=0.0525D$ and $t_3$ from the interval [$0.00625D, 0.2D$] with each curve increasing $t_3$ by a factor of 2. Inset shows the temperature dependence of $G_{12\mathrm{DC}}(\omega)$ DC conductance for the case when $t_1=t_2=0.05D$, $t_3=0.0125D$ with a non-monotonic behavior that is a signature of crossover between two- and three-terminal teleportation. The curve is a fit of $T^{2/3}$ dependence.}
\end{figure*}

\begin{figure}
\includegraphics[width=0.4\textwidth]{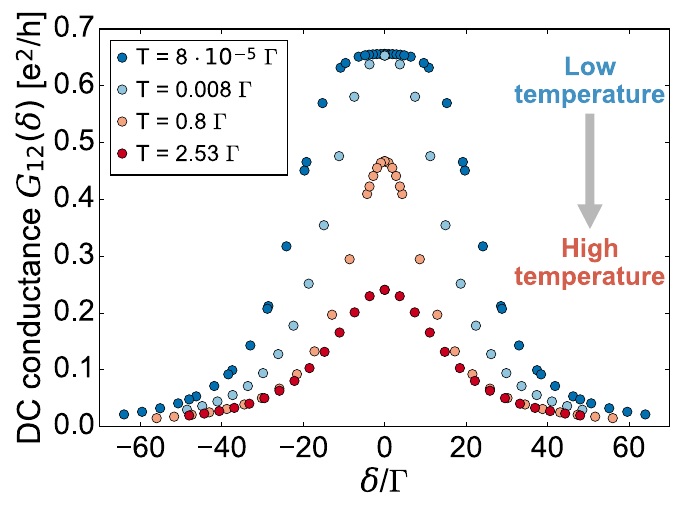}
\caption{\label{fig:charge_degeneracy}(Color online)  $G_{12\mathrm{DC}}(\delta)$ DC conductance away from charge degeneracy point for several temperatures. For low temperatures the top of the curve becomes flattened at the value of 2/3 $e^2/h$, which implies robustness against charge detuning.}
\end{figure}

We begin our analysis of the model with the first property that distinguishes it from the topological Kondo regime studied previously, namely the temperature of the transition from local moment fixed point to the non-Fermi liquid fixed point. The dependence of the transition temperature $T^*$ on the lead coupling parameter can be established in more detail by analyzing the flow of the entropy of the impurities on the island to the non-Fermi liquid fixed point, which is shown in Fig. \ref{fig:entropy}. The entropy values flow from the local moment fixed point with $S_{\mathrm{imp}}(T) = \ln(4)$ to the non-Fermi liquid fixed point with $S_{\mathrm{imp}}(T) = \ln(\sqrt{3})$. When the temperatures are expressed in the units of the level broadening $\Gamma$, all the entropy curves collapse into one universal dependence. Now we can define the transition temperature $T^*$ by numerically solving the equation $S_{\mathrm{imp}}(T^*) = (\ln(4) + \ln(\sqrt{3}))/2$ and plot it as a function of the level broadening (inset of Fig. \ref{fig:entropy}). The curve on which the $T^*$ points lie is defined as $T^* = c \Gamma$, where $c \approx 3.60$ is a constant coefficient. Since there is a direct relation between $T^*$ and $\Gamma$, one can estimate the transition temperature by comparing it with experimentally measured values of level broadening, which are of the order of tens to hundreds $\mu$eV. Such values translate to a temperature of about few K. To contrast this with previous proposals, in Fig. \ref{fig:entropy}(b) we show the comparison between the crossover temperatures $T^*$ for our model and the model in the topological Kondo regime (details of the model in the Appendix B) in the fully isotropic case ($t=t_1=t_2=t_3$). Even for large tunnel couplings $T^*$ at charge degeneracy point is at least 3 orders of magnitude higher than in the Kondo regime. Moreover, the Kondo temperature drops sharply with decreasing couplings ($T_K \sim (\rho t)^2 \exp(-1/(2 \rho t))$), while at charge degeneracy point $T^* \sim t^2$, which can lead to a much easier experimental observation of multi-terminal teleportation. Furthermore, it would be possible to directly measure the dependence of the $T^*$ on the tunnel couplings by tuning them using external gates.

Next, we move onto computing the transport properties of the three-terminal Majorana island. We start by analyzing the results at the charge degeneracy point (when $\delta = 0$). In Fig. \ref{fig:anisotropic} (a), we show the $G_{12}(\omega)$ AC conductance matrix element in the isotropic case ($t_1=t_2=t_3=0.05D$) for varying temperatures. All the computed curves follow a universal dependence and at low temperature the fractional quantized value of $2/3 e^2/h$ is attained as predicted by the quantum Brownian motion mapping. In Fig. \ref{fig:anisotropic} (b) the temperature dependence of the $G_{12\mathrm{DC}}$ DC conductance is shown. The whole crossover happens over the span of approximately two orders of magnitude in temperature, which means it is much steeper than the crossover studied previously in the Kondo regime. This is another factor that makes the experiment possible - the increase of conductance should start at tens of Kelvins and approach the fractional quantized value for tens of mK. The quantum Brownian motion mapping provides a prediction of a universal power-law temperature correction to conductance at the strong coupling fixed point which has the form:
\begin{equation}
G_{12\mathrm{DC}}(T) = G_{12\mathrm{DC}}(T=0)\left(1 - c'(T/T^*)^{2/3}\right).
\end{equation}
The curve presented in the plot is a fit of the predicted dependence and it correctly describes a significant part of the crossover. This fact together with the high crossover temperature makes experimental verification of this low temperature conductance correction possible.

\begin{figure}
\includegraphics[width=0.4\textwidth]{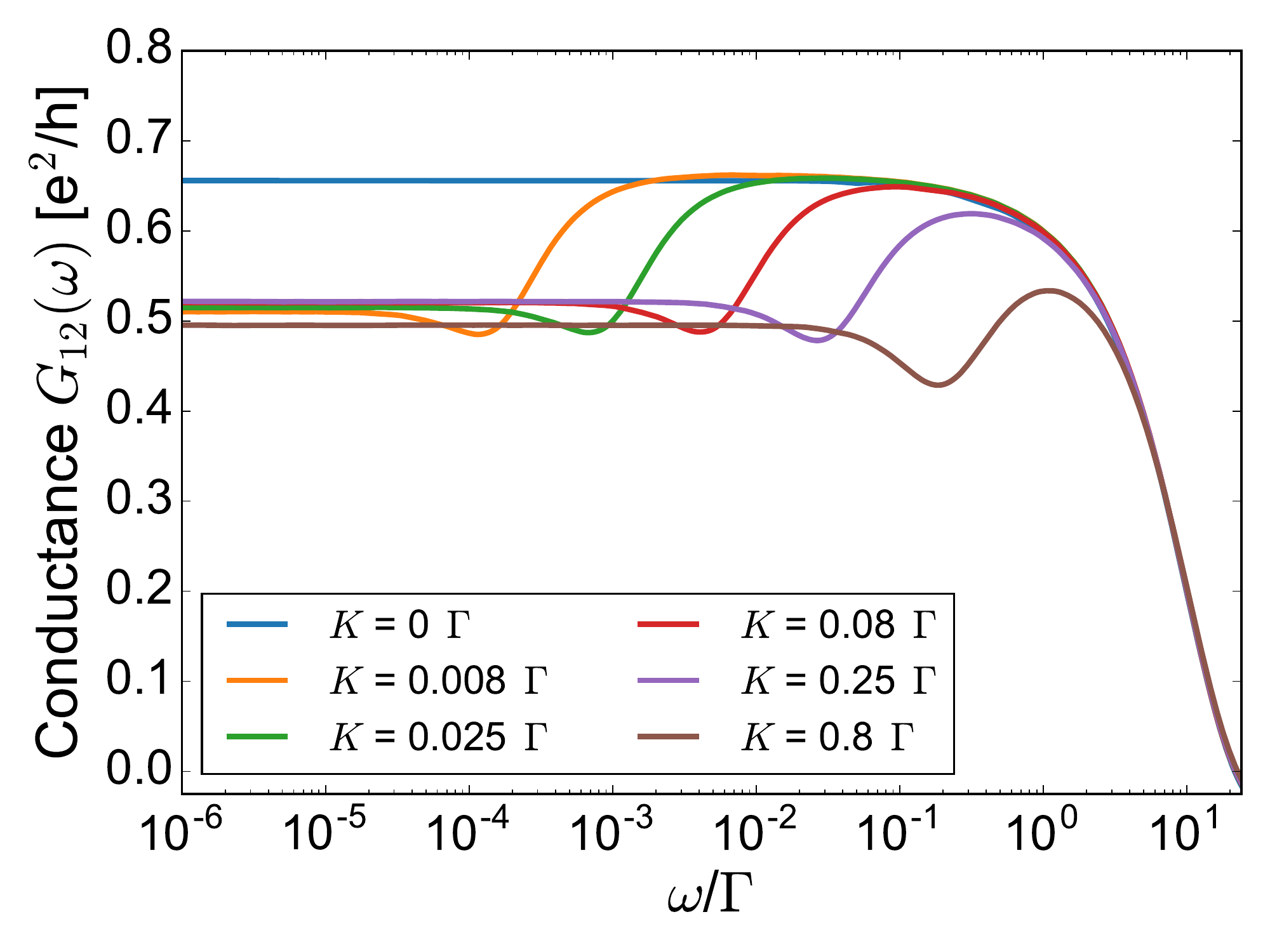}
\caption{\label{fig:G12_hybridization_B_generic}(Color online) $G_{12}(\omega)$ in the isotropic case $t_1=t_2=t_3=0.1D$ for several values of hybridization strength $K_x \neq K_y \neq K_z$ with constant ratio $0.84:1:1.11$ ($K$ being the proportionality constant). The hybridization affects conductance only for very small frequencies (and temperatures), so even a sizable overlap of Majorana states would not preclude experimental observation.}
\end{figure}

\begin{figure}
\includegraphics[width=0.4\textwidth]{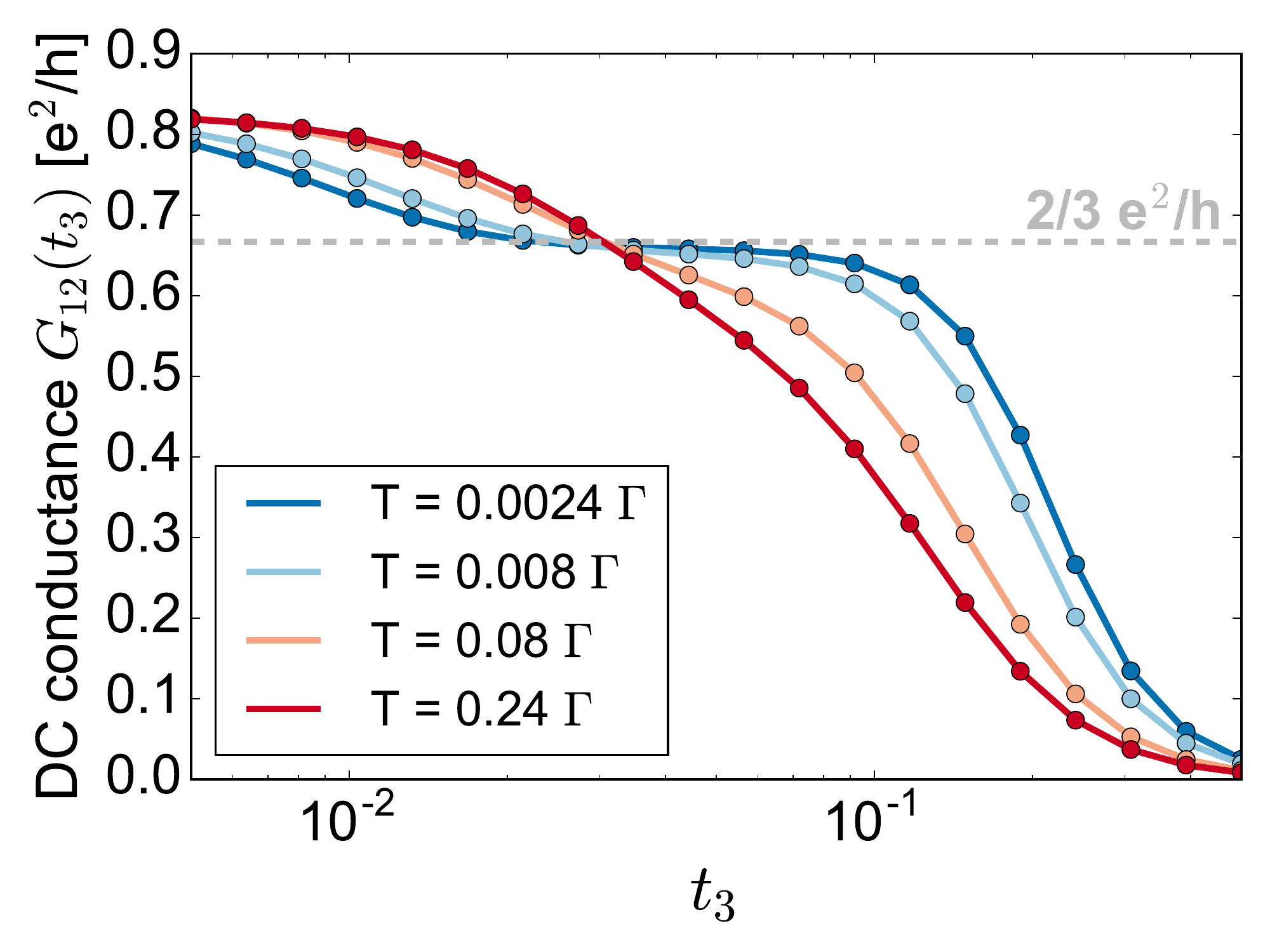}
\caption{\label{fig:G12_DC_J3}(Color online) $G_{12\mathrm{DC}}(t_3)$ DC conductance slightly away from charge degeneracy point ($\delta=0.0035 D$) as a function of tunnel coupling of the third lead for several temperatures. In high temperatures there is a simple transition from 0 to over 0.8 $e^2/h$. When the system is cooled down, a plateau emerges at 2/3 $e^2/h$, signifying transition from 2 to 3 terminal electron teleportation.}
\end{figure}

However, in a real experiment, reaching the exact isotropic case is difficult. Therefore it is important to verify the prediction of robustness to channel coupling asymmetry. In Fig. \ref{fig:anisotropic}(c) we show the results for fully anisotropic set of coupling constants ($t_1 = 0.0475D$, $t_2 = 0.0525D$ and $t_3$ varying in the range $[0.00625D, 0.2D]$ in $T=0$, with each step increasing $t_3$ by a factor of 2. In this case the DC conductance also reaches the value of 2/3 $e^2/h$ independently of the initial value of $t_3$, which is in stark contrast to the s-wave island model. Moreover, in the case of decreasing $t_3$ one can observe a nontrivial crossover between the cases with 2 and 3 leads. For $\omega$ just below $\Gamma$ the value of conductance goes beyond the value of 2/3 and comes close to 1 $e^2/h$, which is the value corresponding to the electron teleportation between only 2 leads. However, going further to lower frequencies decreases conductance and it again attains the fractional quantized value. This behavior is mimicked in the temperature dependence of DC conductance, which is shown in the inset of Fig. \ref{fig:anisotropic}(c). We observe a non-monotonic curve, which first rises above the fractional value for intermediate temperatures, but in the low temperature limit goes back to 2/3 $e^2/h$. The curve is a fit of a $T^{2/3}$ dependence, in this case with a positive coefficient in front of it. This non-monotonic behavior can be used as one of the experimental signatures of crossing between two- and three-terminal teleportation regimes. However, due to the slow decay of conductance back to the fractional value, reaching the low temperature limit may prove to be more difficult.

Another important factor for the experimental verification of our claims is the sensitivity to tuning the system exactly to the charge degeneracy point. In Fig. \ref{fig:charge_degeneracy} we present DC conductance of our system as a function of the energy shift $\delta$ away from the charge degeneracy point for 4 different temperatures. For the lowest temperature, the curve becomes flattened at the top, which corresponds to the conductance value of 2/3 $e^2/h$. This flat top means that even when one moves away from the resonance, the observed conductance would still be equal to the fractional quantized value. For increased temperatures the curves become narrower, but still it is reasonable to expect to observe a non-zero value of conductance even when being away from the charge degeneracy point. Nevertheless, this proves that tuning the system into the vicinity of charge degeneracy point is crucial to observe fractional conductance at the temperatures within the experimental reach.

Finally, we study how the conductance is impacted by introducing Majorana hybridization into our Hamiltonian. Since hybridization is a relevant perturbation, one expects that in low temperatures it will significantly change the behavior of conductance. In Fig. \ref{fig:G12_hybridization_B_generic} we show $G_{12}(\omega)$ in the isotropic case $t_1=t_2=t_3=0.1D$ for several generic values of hybridization strength $K_x \neq K_y \neq K_z$ with constant ratio $0.84:1:1.11$ between the components of $\vec{K}$ (additional results for different values of $\vec{K}$ components are presented in the Appendix B). The conductance rises from 0, reaches value of 2/3 $e^2/h$ and then at lower energy scales changes to some non-universal value. The scale at which the crossover happens depends on the hybridization strength. The most important fact is that even with a sizable magnitude of hybridization, it affects conductance only in the very low temperatures and the fractional quantized conductance still prevails in the range of temperatures available in the experiment. This justifies neglecting the Majorana hybridization in the rest of the calculations.

Having verified the claim of robustness of our results with respect to the tunnel coupling anisotropy, charge degeneracy detuning and showing that hybridization affects the results only at very low temperatures, we propose an experiment which yields a direct signature of the multi-terminal Majorana-assisted electron teleportation in conditions obtainable in a laboratory. In Fig. \ref{fig:G12_DC_J3} we show the DC conductance as a function of the tunnel coupling of the third lead for several different temperatures slightly off the charge degeneracy point to simulate the experimental conditions. In high temperatures, the conductance increases straight to the values close to 1 $e^2/h$ while decreasing the tunnel coupling, as is expected for the electron teleportation between 2 leads. However, as the temperature is lowered a plateau at 2/3 $e^2/h$ emerges and it becomes wider in the process of cooling down the system. Remarkably, the whole shape of the curve changes, the increase of conductance starting for larger tunnel couplings in lower temperatures, which allows to observe the change for a large range of tunnel coupling strengths. This change of conductance curve shape provides a direct evidence of entering the multi-terminal teleportation regime.

\section{Summary}

We have shown that both the conventional and topological superconducting islands at charge degeneracy point are interesting in their own right. By applying bosonization techniques we demonstrated that the multiterminal s-wave superconductor island Hamiltonian maps to the multichannel Kondo problem. For the case of non-interacting leads this means that at low temperatures the system is described by an intermediate coupling fixed point that displays non-Fermi liquid behavior and for which many observables are known from conformal field theory. We supported the mapping by a numerical renormalization group calculation, which gives the residual entropy and conductance consistent with the analytical prediction. The intermediate fixed point is in general unstable to channel coupling asymmetry and so experimental verification would require fine-tuning. On the other hand, due to Luttinger parameter rescaling the topological superconductor island flows to a strong coupling fixed point, which grants robustness to the anisotropy. This conclusion is backed by numerical calculation in which the conductance for $N=3$ leads reaches the value of 2/3 $e^2/h$ independently of the initial tunnel couplings. Moreover, the crossover to non-Fermi liquid fixed point happens at experimentally plausible temperatures, compared to previous studies of topological Kondo effect. Thanks to this robustness, for the topological island we have predicted distinctive experimental signatures of crossover between two- and three-terminal cases:  one is a non-monotonic temperature dependence of DC conductance when coupling of one of the leads is decreased, the other one is the change of the shape of tunnel coupling strength dependence of DC conductance with a plateau emerging at 2/3 $e^2/h$ while decreasing the temperature. As the experimental control of the hybrid semiconductor-superconductor structures sees rapid progress, our predictions may soon be verified in the laboratory.

\begin{acknowledgments}
We thank Karen Michaeli, Aviad Landau and Eran Sela for a previous collaboration on multi-terminal Majorana island. The NRG simulations have been performed using NRG Ljubljana code. This work was supported by the DOE Office of Basic Energy Sciences, Division of Materials Sciences and Engineering under award de-sc0010526.
\end{acknowledgments}

\appendix
\section{Conductance calculation details}

\begin{figure*}
\includegraphics[width=0.99\textwidth]{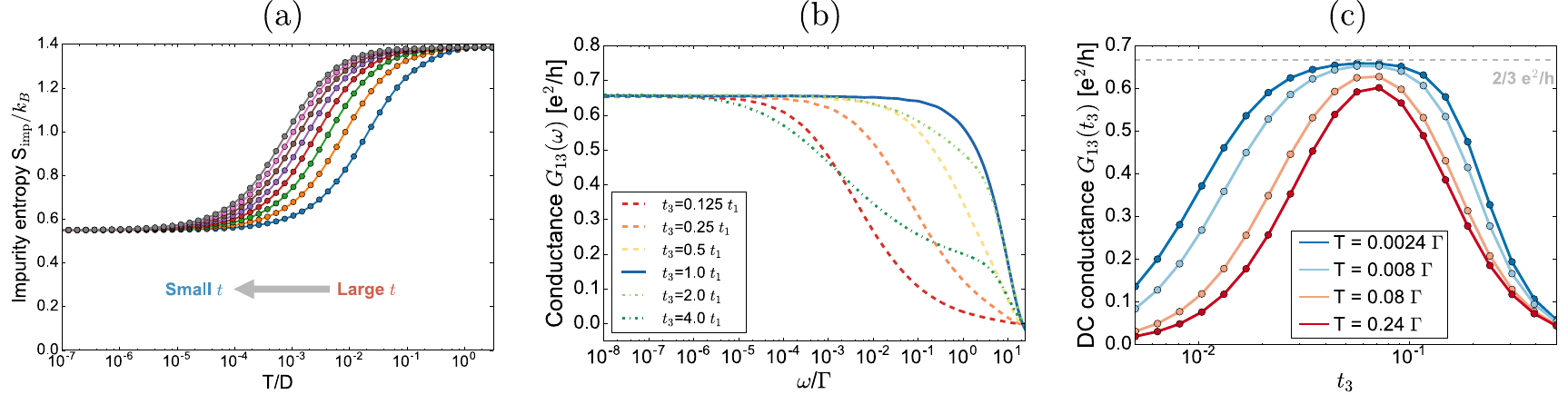}
\caption{\label{fig:B1}(Color online) (a) Entropy of the island's impurity $S_{\mathrm{imp}}(T)$ for temperatures expressing in the units of lead bandwidth $D$ for several tunneling coupling strengths. Entropy flows from $S_\mathrm{imp}(T)=\ln(4)$ for high temperatures to the non-Fermi liquid fixed point with $S_\mathrm{imp}(T)=\ln(\sqrt{3})$. (b) $G_{13}(\omega)$ conductance matrix element in fully anisotropic case $t_1=0.095D$, $t_2=0.105D$ and $t_3$ from the interval [$0.125 t_1, 4t_1$] with each curve increasing $t_3$ by a factor of 2. (c) $G_{13\mathrm{DC}}(t_3)$ DC conductance slightly away from charge degeneracy point ($\Delta_g=0.0035$) as a function of tunnel coupling of the third lead for several temperatures. The conductance has a nonmonotonic dependence, peaked at $t_3 \approx 0.05D = t_1 = t_2$. As the system is cooled down, the value at the peak increases until it reaches 2/3 $e^2/h$. Further lowering the temperature develops a plateau at this value.}
\end{figure*}

\begin{figure*}
\includegraphics[width=0.99\textwidth]{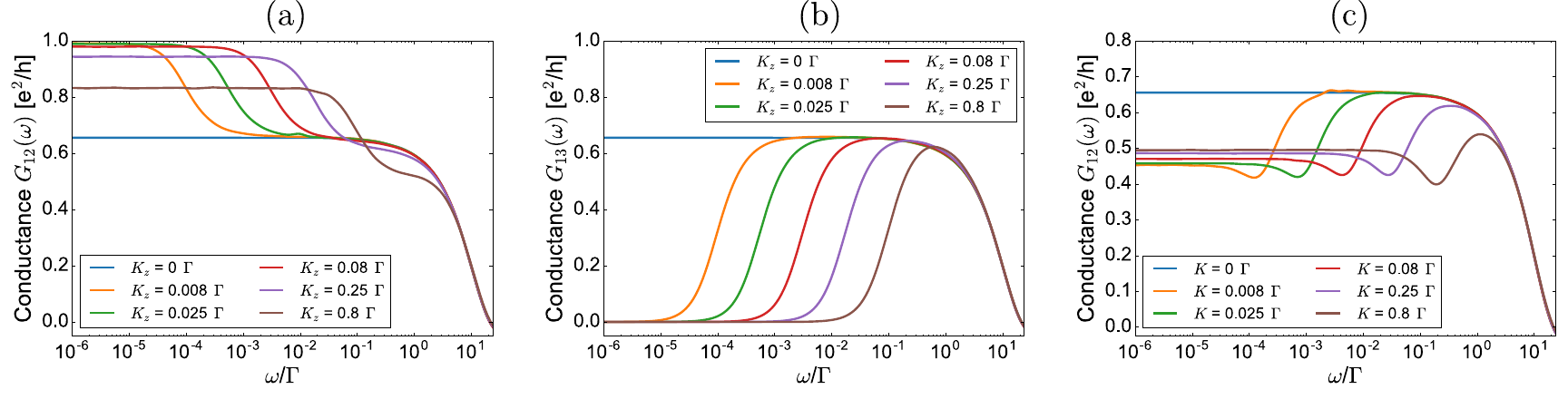}
\caption{\label{fig:B2}(Color online) Conductance matrix element (a)  $G_{12}(\omega)$ and (b) $G_{13}(\omega)$ when Majorana hybridization term is added to the Hamiltonian with $K_x = K_y = 0$, $K_z \neq 0$. (c) $G_{12}(\omega)$ conductance matrix element when Majorana hybridization term is added to the Hamiltonian with $K_x = K_y = K_z = K$. In this case $G_{12}(\omega) = G_{13}(\omega)$.}
\end{figure*}

We calculate conductance in the framework of linear response theory using Kubo formula. To obtain the final expression for conductance we follow a similar procedure as in Appendix B of Galpin et al. \cite{GalpinConductancefingerprintMajorana2014}. Therefore, we study how a perturbation of the form $H'(t) = \hat{\mathcal{O}}_m f(t) e^{\eta t}$ (with $\eta \rightarrow 0$ to account for turning on the perturbation adiabatically) changes the equilibrium expectation value of an operator $\hat{\mathcal{O}_n}$. We use a standard result in the first order of perturbation theory to express the change by:
\begin{equation}
\label{eq:expt_value_general}
\delta \langle \hat{\mathcal{O}_n}(t) \rangle = - \frac{i}{\hbar} \mathrm{Tr} \int_{-\infty}^t e^{\eta t'} [\hat{\mathcal{O}}_m, \rho_{\mathrm{eq}}] \hat{\mathcal{O}}_n(t-t') f(t') dt'
\end{equation}
where we define $\delta \langle \hat{\mathcal{O}_n}(t) \rangle = \mathrm{Tr}(\rho(t)\hat{\mathcal{O}_n} - \rho_{\mathrm{eq}}\hat{\mathcal{O}}_n)$, $\rho_{\mathrm{eq}} = e^{- \beta H}/Z$ and $\rho(t)$ are the density matrices in equilibrium and in the presence of the perturbation, respectively and $\hat{\mathcal{O}}_n(t-t')$ is defined in the interaction picture:
\begin{equation}
\hat{\mathcal{O}_n}(t-t') = e^{\frac{i}{\hbar}\hat{H}(t-t')}\mathcal{O}_n e^{-\frac{i}{\hbar}\hat{H}(t-t')}
\end{equation}

To obtain conductance using the formula \eqref{eq:expt_value_general} we have to study how current through a lead $I_j$ changes when AC voltage $V_k$ is applied to another lead. Therefore, we make the following substitutions: $\hat{\mathcal{O}}_n\rightarrow I_j = e \langle \dot{N}_j \rangle = e \langle \frac{i}{\hbar} [H, N_j] \rangle$, $\hat{\mathcal{O}_n}\rightarrow N_k$ and $f(t) \rightarrow eV_k \cos(\omega t)$. This leads to a formula for the current present in the perturbed system:

\begin{equation}
\label{eq:expt_value_current}
I_j(t) = - \frac{i e^2 V_k}{\hbar} \mathrm{Tr} \int_{-\infty}^t e^{\eta t'} [N_k, \rho_{\mathrm{eq}}] \dot{N}_j(t-t') \cos(\omega t') dt'
\end{equation}

We change the variable of integration $t'' = t-t'$ and define conductance tensor element $\mathcal{G}_{jk}$ as:
\begin{equation}
\begin{split}
&\mathcal{G}_{jk}(t, \omega) = \frac{\partial I_j}{\partial V_k} = \\
 &= - \frac{i e^2}{\hbar} \mathrm{Tr} \int_{0}^\infty e^{\eta (t-t'')} [N_k, \rho_{\mathrm{eq}}] \dot{N}_j(t'') \cos(\omega (t-t'')) dt''
\end{split}
\end{equation}

To simplify the considerations we focus on the value of conductance at $t=0$. Using the cyclic property of trace we arrive at:

\begin{equation}
\begin{split}
&\mathcal{G}_{jk}(t=0, \omega) = \\ &= - \frac{i e^2}{\hbar} \mathrm{Tr} \int_{0}^\infty e^{-\eta t''} \rho_{\mathrm{eq}} [N_k, \dot{N}_j(t'')] \cos(\omega t'') dt'' = \\ &= - \frac{i e^2}{2 \hbar} \int_{0}^\infty e^{-\eta t''} \langle [N_k, \dot{N}_j(t'')] \rangle (e^{i \omega t''} + e^{-i \omega t''}) dt''
\end{split}
\end{equation}

Now we insert the complete basis of energy states and compute the conductance using Lehmann spectral representation. We finally arrive at:
\begin{equation}
\mathcal{G}_{jk}(\omega) = \frac{e^2}{2\hbar} \left( \sigma_{jk}(\omega) + \sigma_{jk}(-\omega) \right)
\end{equation}
with
\begin{equation}
\label{eq:sigma}
\begin{split}
\sigma_{jk}(\omega) = \frac{1}{Z} \sum_{m,n} \frac{E_n - E_m + \omega - i \eta}{(E_n-E_m+\omega)^2 + \eta^2} \left(e^{-\beta E_m} - e^{-\beta E_n}\right) \\ \times \langle m| N_k |n\rangle \langle n| \dot{N}_j|m\rangle
\end{split}
\end{equation}

During the NRG simulation we compute the imaginary part of $\sigma_{jk}(\omega)$. The real part can be obtained afterwards by performing Kramers-Kronig transformation. The quantity $G_{jk}(\omega)$ we show in the figures is:
\begin{equation}
G_{jk}(\omega) = \mathrm{Im}\, \mathcal{G}_{jk}(\omega) = \frac{e^2}{h} \pi \left( \mathrm{Im}\, \sigma_{jk}(\omega) + \mathrm{Im}\, \sigma_{jk}(-\omega) \right)
\end{equation}

The advantage of the method presented above is apparent when one calculates the DC conductance as a limit $\omega \rightarrow 0$. The usual approach is to compute:
\begin{equation}
G_{jkDC} = - 2\pi \lim_{\omega \rightarrow 0} \frac{\mathrm{Im}\, K(\omega)}{\omega}
\end{equation}
where
\begin{equation}
K(\omega) = - \frac{i}{\hbar} \int_0^\infty e^{i(\omega + i \eta) t} \langle [\dot{N}_j, \dot{N}_k(t)] \rangle dt
\end{equation}

This approach involves calculation of a limit of a ratio of two very small quantities, which may prove to be unreliable numerically. The trade-off of the method we used is that it requires computation of global operators $N_j$, which depend not only on the impurity, but also on the sites of the Wilson chains.

\section{Additional Majorana island NRG results}
In this section we present additional NRG simulation results. We begin with the details of the model we are comparing our results to. The model describes the topological superconductor island in the Coulomb valley regime. The Hamiltonian in this case is \cite{GalpinConductancefingerprintMajorana2014}:
\begin{equation}
\begin{split}
\hat{H} &= \hat{H}_{\mathrm{leads}} + \frac{t_1}{\sqrt{2}} (\sigma^+ \psi^\dagger_0 \psi_1 + \sigma^- \psi^\dagger_1 \psi_0)\, + \\ &+ \frac{t_2}{\sqrt{2}} (\sigma^+ \psi^\dagger_{-1} \psi_0 + \sigma^- \psi^\dagger_0 \psi_{-1})\, + t_3 \sigma_z (\psi^\dagger_1 \psi_1 - \psi^\dagger_{-1} \psi_{-1})
\end{split}
\end{equation}
where $\psi_j$ are annihilation operators at the ends of the three spinless leads and $\sigma$ are the spin operators of the impurity formed on the island. The Hamiltonian is obtained by considering virtual transitions between leads in second order perturbation theory. This results in much stronger crossover energy scale dependence on the tunnel couplings and is one of the reasons for many orders of magnitude of difference between the transition temperature at charge degeneracy point and in the Kondo regime.

In Fig. \ref{fig:B1}(a) we show the entropy $S_\mathrm{imp}(T)$ curves with temperature expressed in units of the lead bandwidth $D$, before collapsing all of them onto one curve as shown in the main text. In Fig. \ref{fig:B1}(b) we show the conductance matrix element $G_{13}(\omega)$ for several values of the tunnel coupling of the third lead $t_3$ (complementary plot to Fig.~\ref{fig:anisotropic}(c) from the main text). In this case one can also observe the transition to two-terminal teleportation: for $t_3 < t_1, t_2$ the conductance reaches the fractional quantized value of 2/3 $e^2/h$ only for very low frequencies and analogously, very low temperatures. For higher temperatures the conductance is essentially 0 (in the same regime $G_{12}(\omega)$ is close to 1 $e^2/h$). In Fig. \ref{fig:B1}(c) we show the DC conductance $G_{13\mathrm{DC}}(t_3)$ for several different temperatures (complementary plot to Fig.~\ref{fig:charge_degeneracy}(a) from the main text). In this case, the conductance forms a peak with the maximum for $t_3$ close to the isotropic case. When the temperature is decreased, at first the height of the peak increases, but when it reaches 2/3 $e^2/h$ the increase stops and instead a plateau is developed. This can also serve as an experimental signature of multi-terminal electron teleportation.

In Figs. \ref{fig:B2}(a) and (b) we present the results of calculations with hybridization term that includes only the $z$ component of $\vec{K}$. Since $K_z \sigma_z \sim i\gamma_1 \gamma_2$, this term connects Majorana states coupled to leads 1 and 2, effectively decoupling the third lead. This in turn gives 1 $e^2/h$ conductance in very low temperatures, the same as in the case of two lead electron teleportation. At the same time, conductance $G_{13}(\omega)$ drops to 0 as a result of this decoupling. When the components of $\vec{K}$ are all equal, the conductance is the same both in case of $G_{12}(\omega)$ and $G_{13}(\omega)$ and is similarly equal to 2/3 $e^2/h$ before decreasing to some non-universal value between 0 and 2/3.

\bibliography{multichannel_charge_kondo}

\end{document}